# Attractive energy and entropy or particle size: the yin and yang of physical and biological science[1]

Douglas Henderson

Department of Chemistry and Biochemistry, Brigham Young University, Provo UT 84602, USA

[1] This article is an expanded version of invited talks given at the Society for Industrial and Applied Mathematics (SIAM) ``Symposium on multi-scale nonlinear problems in biological systems: experiments, numerics, and theory" , Snowbird, UT, USA in May, 2007 and the Johann Radon Institute for Computational and Applied Mathematics (RICAM) ``Semester on quantitative biology analyzed by mathematical methods", Linz, Austria, October, 2007.

# Abstract

It is well known that equilibrium in a thermodynamic system results from a competition or balance between lowering the energy and increasing the entropy, or at least the product of the temperature and entropy. This is remarkably similar to the Taoist concept of yin, a downward influence, and yang, an upward influence, where harmony is established by balancing yin and yang. Entropy is due to structure, which is largely determined by core repulsions or particle size whereas energy is largely determined by longer range attractive interactions. Here, this balance between energy and entropy or particle size is traced through the theory of simple fluids, beginning with Andrews and van der Waals, the subsequent developments of perturbation theory, theories of correlation functions that are based on the Ornstein-Zernike relation and the mean spherical approximation, electrolytes, and recent work on ion channels in biological membranes, where the competition between energy and size gives an intuitively attractive explanation of the selectivity of cation channels. Simulations of complex systems, including proteins in aqueous solution, should be studied to determine the extent to which these concepts are useful for such situations.

# 1 Introduction

In a mechanical system, the equilibrium state is the state of lowest energy, $E$. For example, a ball will sit at the bottom of an energy well or bowl. However, what happens if the table on which the bowl is placed shakes, perhaps due to an earthquake (a common occurrence in the western US)? The ball will, on average, be in a higher energy state and no longer always be at the bottom of the bowl. The counterpart of the earthquake in a thermodynamic system is the temperature, $T$. Equilibrium in a thermodynamic system is the state of lowest free energy, which in a $TVN$ system ($V$ is the volume, and $N$ is the number of particles) is $A = E - TS$, where $A$ and $S$ are the Helmholtz free energy and entropy. The Helmholtz function and the entropy are related by

$$S = -\frac{\partial A}{\partial T}. \tag{1}$$

Thus, thermodynamic equilibrium is established by a competition or balance between decreasing $E$ and increasing $S$ (or at least $TS$). This balance is remarkably similar to the Toaist concept of yin and yang, yin being a downward influence and yang being an upward influence, where harmony, or balance, in life is achieved through balancing yin and yang.

The concept of entropy was established by Clausius and others in the mid nineteenth century and its tendency to increase was understood by them. However, our modern understanding of the relation of entropy to the disorder of a system was established by Boltzmann towards the end of the nineteenth century. The genius of Boltzmann was not realized at that time and perhaps is not fully realized today. Boltzmann was criticized because his ideas contradicted deterministic Newtonian mechanics. His work anticipated non-deterministic mechanics, which is only now being broadly developed. Because of the revolutionary and far reaching character of Boltzmann's work, he is a first tier scientist, similar to Newton, Maxwell, and Einstein. It is sad that Boltzmann became depressed because of the criticism of lesser, but politically more powerful, scientists and ended his life.

# 2 van der Waals

This article will focus on liquids. Firstly, this is because I was attracted to the study of liquids because, when I was young, it was believed that, in contrast to gases and solids, there was no useful theory of liquids. Indeed, as a graduate student I recall reading an essay by Joe Hirschfelder in which the lack of a theory of liquids was stated to be a major bottleneck in physical science. Secondly, biophysics has emerged as an exciting field for physical scientists. There may be exceptions, but most biology occurs in liquids (water or an aqueous solution) or at the interface of a liquid and a solid.

The theory of liquids started with the experiments of Andrews who in the nineteenth century established the *continuity of states* between gases and liquids. Figures 1 and 2 show the coexistence of a typical liquid and gas. The gas exists at low densities and the liquid exists at higher densities. The density is $\rho = N/V$. If at low (but not too low) temperatures the gas is compressed, it condenses into a liquid. Similarily, the liquid can vaporize to a gas. The vapor-liquid phase transition is first order. However, as the temperature is increased one arrives at the critical temperature, $T_c$, where the gas and liquid phases coincide and the transition becomes second order. The region in Figs. 1 and 2 with $T$ below $T_c$ (but above the triple point temperature, $T_t$) and above the critical density, $\rho_c$, (but below the densities of the solid-fluid transition) is generally called the liquid state and the remaining region (again below the densities for solid-fluid coexistence) is generally called the gaseous state. This is a poor terminology. A gas will occupy all the available volume whereas a liquid will not. A liquid is characterized by a free surface. Thus, only the high density fluid along the liquid-vapor equilibrium curve is clearly a liquid. It would be best to reserve the terms liquid and vapor for the high and low density fluids that are in liquid-vapor equilibrium, respectively, and regard the fluid in the remaining region, outside the solid state, as a fluid or gas. However, this careless terminology is so ingrained that it is likely fruitless, perhaps even pedantic, to attempt a reformation. In any case, there is a continuity of states. By increasing $T$ then a compression followed by a lowering of the temperature, a vapor can pass continuously to a liquid and *vice versa*. Of course, we are all aware that the phase diagram in Figs. 1 and 2 are for a simple fluid. Water is more complex. At low temperatures, liquid water is more dense than ice. Indeed, although this is annoying if an automobile radiator freezes, aquatic life would not be possible in colder climates if this were not true. Still, for our discussion this is a detail. The above discussion is applicable to water/water vapor equilibrium.

Presumably, van der Waals (vdW) was aware of Andrews work. In any case, vdW thought of the intermolecular interactions between fluid molecules as being attractive at long range and repulsive at short range. This is illustrated, for spherically symmetric molecules, in Fig. 3 with the Yukawa intermolecular pair interaction,

$$u(R) = \begin{cases} \infty, & R < \sigma \\ -\dfrac{\varepsilon\sigma}{R}\exp[-\dfrac{z(R-\sigma)}{\sigma}], & R > \sigma, \end{cases} \tag{2}$$

where $R$ is the separation of the centers of a pair of molecules, $\sigma$ is the molecular diameter, and $\varepsilon$ is the strength of the energetic interaction at contact. In the case of the Yukawa potential, this is also the minimum of the pair potential. The parameter $z$ determines the range of the potential. The Yukawa potential is only qualitatively reasonable as a representation of the interparticle potential. However, the thermodynamic properties of a simple fluid are determined largely by the area of the potential curve that can be given accurately by adjusting $z$. The value $z = 1.8$ is appropriate for most liquids (Henderson *et al*, 1978)

Starting with the ideal gas equation,

$$pV = NkT, \tag{3}$$

or equivalently,

$$\frac{A_0}{NkT} = 3\ln\lambda - 1 + \ln N - \ln V, \tag{4}$$

where $k$ is the Boltzmann constant (the gas constant per molecule) and $\lambda = h/(2\pi mkT)^{1/2}$ with $h$ and $m$ being Planck's constant and the molecular mass, vdW approximated the contribution of the repulsive term, due to $\sigma$, as resulting in a reduced available volume (or free volume), $V - Nb$. Assuming that the ideal gas equation is still valid,

$$p_0 = \frac{NkT}{V - Nb} \tag{5}$$

and

$$\frac{A_0}{NkT} = 3\ln\lambda - 1 + \ln N - \ln(V - Nb). \tag{6}$$

The subscript 0 is used to emphasize that $p_0$ and $A_0$ are the hard sphere contributions to $p$ and $A$. Van de Waals regarded the contribution of the attractive forces as an *internal pressure* that held the fluid together (after all, liquids can exist at atmospheric pressures that are quite small). Since the internal pressure would surely vanish at low density, there is every reason to expect that a reasonable approximation would be

$$\frac{A}{NkT} = \frac{A_{id}}{NkT} - \ln(1 - \rho b) - \frac{\rho a}{kT}, \tag{7}$$

where the subscript *id* indicates that the quantity is the ideal gas term. Since

$$p = -\frac{\partial A}{\partial V}, \tag{8}$$

the famous van der Waals equation,

$$p = \frac{NkT}{V - Nb} - \frac{N^2 a}{V^2}, \tag{9}$$

results. It will be helpful to write this as

$$p = p_0 - \frac{N^2 a}{V^2}. \tag{10}$$

The entropy is given by Eq. (1). Differentiation with respect to $T$ yields

$$\frac{S}{Nk} = \frac{S_{id}}{Nk} + \ln(1 - \rho b) \tag{11}$$

and

$$\frac{E}{NkT} = \frac{E_{id}}{NkT} - \frac{Na}{VkT}. \tag{12}$$

The parameters $a$ and $b$ can be regarded as adjustable parameters that can be fit to experiment. Although giving insight into fluids, it was immediately realized that the vdW equation was imperfect. As the critical point is a single point and the value of $pV/NkT$ at the critical point is a single number. Independently of the choice for $a$ and $b$, its value at the critical point is 3/8=0.375 in the vdW theory, whereas the experimental value is considerably lower (around 0.29). In retrospect, too much emphasis was placed on the value of $pV/NkT$ at the critical point because the thermodynamic functions are now known to be singular at the critical point. The study of the critical point is really a separate field. In any case, as is seen in Table 1, the vdW theory is not too unreasonable for a gas at low densities but quite deficient for a gas or liquid at high densities. The vdW pressures are too large, often by an order of magnitude compared to the experimental results (Hsu and McKetta, 1964).

One notable feature of the vdW equation is that it predicts a *law of corresponding states*. That is,

$$p / p_c = f(V / V_c, T / T_c), \tag{13}$$

where $f$ is some function. Thus, the equations of state of various similar fluids can be scaled so that they become one equation. This is a good approximation for many liquids although such effects as molecular shape prevent such a simple relation from being exact. The vdW expression for $f$ is in error but the concept of corresponding states is not.

Improvements were attempted. For example, Berthelot suggested

$$p = \frac{NkT}{V - Nb} - \frac{N^2 a}{V^2 T} \tag{14}$$

and Dieterici suggested

$$p = \frac{NkT}{V - Nb} \exp\left[ -\frac{Na}{VkT} \right]. \tag{15}$$

The Dietericic equation yields a better value for $pV / NkT$ at the critical point. However, both the Berthelot and Dieterici equations fail at high densities. Because of this, for a century the vdW theory was regarded, incorrectly, as having only pedagogical interest.

**Table 1** Equation of state of methyl chloride ($T_c = 143.1^\circ$ C, $p_c = 65.919$ atm, $V_c = 2.755$ cm$^3$/gm)

| $T$ ($^\circ$C) | $V$ (cm$^3$/gm) | $p$ (expt1) (atm) | $p$ (vdW) (atm) |
|---|---|---|---|
| 125 | 88.266 | 6.975 | 7.063 |
| | 27.774 | 20.049 | 20.629 |
| | 16.333 | 30.664 | 32.017 |
| | 11.070 | 39.817 | 42.215 |
| | 7.204 | 50.015 | 52.848 |
| | | | |
| 143.7 | 67.069 | 9.567 | 9.632 |
| | 18.099 | 30.398 | 31.288 |
| | 8.321 | 51.077 | 53.818 |
| | 1.939 | 69.954 | 251.92 |
| | 1.610 | 101.510 | 936.55 |
| | 1.400 | 206.85 | 4265. |

| | | | | | | | | |
|---|---|---|---|---|---|---|---|---|

Even as late as the middle of the twentieth century, equations similar in form to that of vdW were suggested. For example, in the mid twentieth century Redlich and Kwong proposed the equation of state

$$p = \frac{NkT}{V - Nb} - \frac{N^2 a}{T^{1/2} V(V + Nb)}. \tag{16}$$

All of these attempted improvements retain the vdW free volume approximation, $V - Nb$. Investigators were directing their attention to the wrong term. The problem with the vdW equation and its decendents was not the energy term involving $a$, but with the overly simplified free volume. The hard core term is treated crudely. This can be seen by returning to the hard sphere fluid, which would be described by Eq. (5) if the vdW equation were valid. Expanding Eq. (5) in powers of the density, $\rho = N/V$ yields

$$\frac{p_0 V}{NkT} = 1 + \rho b + (\rho b)^2 + (\rho b)^3 + \cdots. \tag{17}$$

This is a one dimensional approximation that neglects the fact that in higher dimensions, molecules can go around each other. The pressure can be expanded in a series in $\rho$. In three dimensions,

$$\frac{p_0 V}{NkT} = 1 + \rho b + \frac{5}{8}(\rho b)^2 + 0.2869(\rho b)^3 + \cdots, \tag{18}$$

with $b = 2\pi\sigma^3/3$, where $\sigma$ is the hard sphere diameter.

The vdW free volume grossly overestimates the hard sphere contribution to the pressure. This can be seen in Fig. 4. Since Eq. (18) was known in the nineteenth century, it seems quite amazing that the vdW free volume was retained for so long. A simple alternative,

$$\frac{p_0 V}{NkT} = \frac{1}{(1-\eta)^4}, \tag{19}$$

$\eta = \pi\rho\sigma^3/6$ can be constructed easily (Guggenheim, 1969). Equation (19) was known in the nineteenth century and was rediscovered about forty years ago. The vdW expression correctly reproduces only the first correction, called the second virial coefficient, to the ideal gas expression for $pV/NkT$, whereas Eq. (19) reproduces the second and third virial coefficients correctly and gives a reasonable estimate of the higher virial coefficients. An even better approximation is the CS equation (Carnahan and Starling, 1965),

$$\frac{p_0 V}{NkT} = \frac{1 + \eta + \eta^2 - \eta^3}{(1-\eta)^3}. \tag{20}$$

The CS equation also yields the correct second and third hard sphere virial coefficients and gives slightly better estimates for the higher order virial coefficients. In view of this comparison with the exact hard sphere virial coefficients, it is no surprise that, as is seen in Fig. 4, where a comparison with simulation values for $p_0$ (Barker and Henderson, 1971) is shown, the vdW hard sphere equation of state gives values for the pressure that are much too large. In contrast, the results of Eqs. (19) and (20) are much better. The CS result is very accurate. Much better results can be obtained using Eq. (10) but with some better expression for $p_0$, say the CS expression (Longuet-Higgins and Widom, 1965). Thus, we could regard

$$p = \rho kT \frac{1+\eta+\eta^2-\eta^3}{(1-\eta)^3} - \rho^2 a \tag{21}$$

as an improved or modified vdW equation.

The reason the vdW expression for $p_0$ was retained so long was the fact that, with the vdW free volume approximation, the vdW equation of state is a cubic equation that can be solved analytically. Before the advent of digital computers and, more especially, inexpensive personal computers, computations based upon the more accurate formulae were very time consuming. Perhaps the use of the vdW free volume is understandable for practical computations. However, it is very surprising that research scientists did not realize the the vdW theory was the basis of the desired theory of liquids and not just something for professors to mention in their lectures.

To cast this in terms of the aforementioned balance or competition between energy and entropy, note that in the vdW equation of state and in many of its extensions, the attractive forces, through the parameter, $a$, contribute *exclusively* to $E$ and the repulsive forces (or particle size), through the parameter $b$, contribute *exclusively* to the entropy. This separation of energetic and entropic terms is absolute in the vdW theory. It is not absolute in more accurate theories that will now be considered but is still a good approximation.

## 3 Perturbation theory

Perturbation theory provides the formalism whereby the vdW can be made systematic and extensions developed. Perturbation theory (at least to first order) was developed by Zwanzig and Longuet-Higgins but was regarded as a high temperature theory that was useful only for gases. It was BH (Barker and Henderson, 1967a) who first realized that perturbation theory was useful for liquids and then estimated and calculated the second order term.

Considering first a pair potential with a hard core such as, for example. the Yukawa potential, the Helmholtz function can be expanded in powers of $\beta\varepsilon$, where $\beta = 1/kT$. The result is

$$A = A_0 + \sum_{n=1}^{\infty} (\beta\varepsilon)^n A_n, \tag{22}$$

where $A_0$ is, as before, the Helmholtz function of the *unperturbed* hard sphere system and the $A_n$, for $n > 0$, give the effect of the perturbing attractive potential. The CS expression for $A_0$ is

$$\frac{A_0}{NkT} = \frac{A_{id}}{NkT} + \eta \frac{4-3\eta}{(1-\eta)^2}. \tag{23}$$

Of course, such an expansion might turn out to have only a small region of convergence.

However, all evidence indicates that this expansion is convergent for liquid temperatures between the critical and triple point temperatures. The first order term is easily obtained [7] and is given by

$$\frac{A_1}{NkT} = 2\pi\rho\int_{\sigma}^{\infty}\left\{\frac{u(R)}{\varepsilon}\right\}g_0(R)R^2dR, \tag{24}$$

where $\varepsilon$ is the depth of the pair potential, $u(R)$. The function $g(R)$ is the *pair correlation function*, a *radial distribution function* (RDF) in the case of spherical particles, that gives the probability of finding a particle a distance $R$ from a central particle, normalized to unity at large $R$. The subscript 0 indicates that $g_0(R)$ is the RDF for the *unperturbed* hard sphere fluid. A typical hard sphere RDF is shown in Fig. 5. The central molecule is surrounded by about 12 nearest `neighbors' (nn) at a distance of about $\sigma$. In a solid the nearest neighbor distance (nnd) would be almost exactly $\sigma$ and the nn peak would be nearly a delta function. However, in a fluid, the nnd is diffuse. The nn's tend to exclude molecules further out and produce a minumum. The neighbors of neighbors produce further maxima and minima of diminishing strength. The asymptotic value, as $R$ becomes large, of a RDF is 1 because the particles are uncorrelated at large $R$.

The term $A_1$ is plotted in Fig. 6 for the Yukawa potential. This curve is typical of the $A_1$ that is obtained using other $u(R)$. At low densities, the higher order $A_n$ have the form (Barker and Henderson, 1967a),

$$\frac{A_n}{NkT} = \frac{(-1)^{n-1}}{n}2\pi\rho\int_{\sigma}^{\infty}\left\{\frac{u(R)}{\varepsilon}\right\}^n g_0(R)R^2dR + \cdots. \tag{25}$$

The remaining terms are higher order in $\rho$ and involve higher order correlation functions. Barker and Henderson argued that these higher order terms are small, especially at high densities and that the second order term, $A_2/NkT$ is, to a good approximation, proportional to the compressiblity of the unperturbed hard sphere fluid. They called this a *compressiblity approximation*. A simple version of a compressiblity approximation for $A_2$ is

$$\frac{A_2}{NkT} = -\pi\rho kT\left(\frac{\partial\rho}{\partial p_0}\right)\int_{\sigma}^{\infty}\left\{\frac{u(R)}{\varepsilon}\right\}^2 g_0(R)R^2dR. \tag{26}$$

The subscript 0 in the compressiblity indicates that the unperturbed hard sphere compressiblity is used. Using the CS equation of state,

$$kT\left(\frac{\partial\rho}{\partial p_0}\right) = \frac{(1-\eta)^4}{1-2\eta-2\eta^2-10\eta^3+\eta^4}. \tag{27}$$

The results obtained from Eqs. (26) and (27) using the Yukawa potential are very similar to the results shown in Fig. 7 for $A_2$ that were obtained using more refined methods. Although the compressiblity approximation is simple, it is quite accurate and subsequent approximations for $A_2$ are not significantly better. In any case, independently of its accuracy, the compressiblity approximation gives insight to the features of $A_2$.

Again, $A_1$ contributes to the energy but not the entropy. In first order, the entropy of the fluid (or liquid) is that of a hard sphere fluid. The first order term is a background energy, determined from the structure of a hard sphere fluid that produces the internal pressure and provides a well in which the molecules move as if they were hard spheres. That is, to first order in $A$, the particles are distributed as if they were hard spheres. The task is to determine $g_0(R)$. This can be done by computer simulation or by means of some theory. The higher order $A_n$ contribute to both the entropy and energy. The division into terms that contribute only to the entropy or only to the energy is no longer true if higher order terms are included.

One simple approximation is $g_0(R)=1$ for $R>\sigma$. This yields

$$\frac{A_1}{NkT}=-\rho a, \tag{28}$$

with

$$a=-2\pi\int_{\sigma}^{\infty}u(R)R^2dR. \tag{29}$$

In the case of the Yukawa potential this gives

$$a=2\varepsilon\sigma^3\frac{z+1}{z^2}. \tag{30}$$

Since Eq. (28) is just the last term in the vdW result, Eq. (7), we can refer to $g_0(R)=1$ as the vdW approximation for $A_1$.

The higher order $A_n$ are increasingly difficult to compute. For example, the expression for $A_2$ involves two, three, and four particle correlation functions. Barker and Henderson determined $A_2$ by MC simulation as well as by a compressiblity approximation. An approximate integral equation that permits the calculation of the $A_n$ will be discussed in the next section. For the moment, the point to be made is that for a hard core potential, such as the Yukawa potential, first order perturbation theory predicts that the entropy is determined by the hard core repulsive potential and the energy change is due to the attractive part of the pair potential. Equilibrium is established by a balance or competition between excluded volume or space and attractive energy.

## 4 Theories based on the Ornstein-Zernike relation

Integral equations provide an approximate method of calculating $g(R)$ and thermodynamics but also the higher order $A_n$. Most modern integral equation theories are based on the *Ornstein-Zernike* (OZ) relation. To obtain this relation we observe first that subtracting its asymptotic value from $g(R)$ yields a correlation function, called the *total correlation function*, $h(R)=g(R)$-1. This function is not long ranged in the sense of the Coulomb potential but its range can extend over several molecular diameters.

It is useful to separate $h(R)$ into a *direct correlation function* (DCF), $c(R)$, and an indirect term. At infinite dilution, we would expect $h(R)$ and $c(R)$ would be equal since there are only pairwise encounters at such low densities. However as the density is increased, triplet encounters may occur. Molecules 1 and 2 may be correlated directly but also indirectly with a third molecule that is directly correlated with molecule 1 and molecule 2. Thus,

$$h(R_{12}) = c(R_{12}) + \rho\int c(R_{13})c(R_{23})dr_3, \tag{31}$$

where $r_i$ is the position of the center of molecule $i$ and $R_{ij} = |r_i - r_j|$. To keep things simple, only spherically symmetric molecules have been considered. In the case of nonspherical molecules, an integration over angular variables is required. The integral in Eq. (31) is a convolution integral. If the Fourier transform is taken, such a convolution integral becomes the product of the Fourier transforms of the two terms in the integral. Thus, in Fourier space

$$\tilde{h}(k) = \tilde{c}(k) + \rho\tilde{c}^2(k), \tag{32}$$

where $k$ is the Fourier transform variable and not the Boltzmann constant. As soon as this discussion is completed, $k$ will revert to representing the Boltzmann constant.

As the density is increased, the indirect function will involve chains of several molecules. This series is called a *chain sum*. Thus, in Fourier space,

$$\tilde{h}(k) = \tilde{c}(k) + \rho\tilde{c}^2(k) + \rho^2\tilde{c}^3(k) + \cdots \tag{33}$$

or

$$\tilde{h}(k) = \tilde{c}(k)\left(1 + \rho\tilde{c}(k) + \rho^2\tilde{c}^2(k) + \ldots\right), \tag{34}$$

which can be summed to yield

$$\tilde{h}(k) = \frac{\tilde{c}(k)}{1 - \rho\tilde{c}(k)} \tag{35}$$

or, equivalently,

$$\tilde{h}(k) = \tilde{c}(k) + \rho\tilde{h}(k)\tilde{c}(k). \tag{36}$$

Converting back to coordinate space, we have the OZ relation

$$h(R_{12}) = c(R_{12}) + \rho\int h(R_{13})c(R_{23})dr_3. \tag{37}$$

Originally, this relation was obtained to describe light scattering, especially near the critical point. However, during the past half century it has been realized that the OZ relation can be used to develop theories of fluids. By itself, the OZ relation is merely a definition of the DCF and does not give a method of determing $c(R)$ unless $h(r)$ is already known. To yield an approximation, the OZ relation must be coupled with some approximate relation, called a *closure* because the closure yields a closed equation. If the molecules have a hard core, *ie,* $u(R) = \infty$ for $R < \sigma$. Therefore, we have the exact result,

$$h(R) = -1 \tag{38}$$

for $R < \sigma$, where $\sigma$ is the hard core diameter. Thus, for such a fluid it is necessary only to specify the DCF (say) for $R > \sigma$. Here, we consider only one closure, the *mean spherical approximation* (MSA), where it is assumed that

$$c(R) = -\beta u(R), \tag{39}$$

for $R > \sigma$. Equations (38) and (39), together with the OZ relation, permit the determination of $c(R)$ and $g(R) = h(R) + 1$. There is no reason to suppose, *a priori*, that the MSA is particularily accurate. However, *a posteriori*, it turns out to be quite accurate for many applications. Not only that, the MSA has an analytic solution for many systems. The accuracy of a closure is determined by comparison with simulations and experiment. Other criteria, such as the completeness of the set of terms in $c(R)$, have not proven to be very useful in assessing the accuracy of a closure.

The name mean spherical approximation conveys little. The origin of this name comes from the theory of lattice gases and, no doubt, once meant something. In any case, Eq. (39) is exact at large $R$ and Eq. (38) is the exact hard core condition. The MSA can be considered to be a hard core approximation that is correct at large $R$.

It is worth mentioning that for the hard sphere fluid, the MSA consists of Eq. (38) together with

$$c(R) = 0 \tag{40}$$

for $R > 0$. It is interesting to note that this is exactly the *Percus-Yevick* (PY) approximation for a hard sphere fluid. The PY approximation and the MSA are identical for hard spheres. However, they are different approximations for other fluids. Generally speaking, the PY approximation has been found to be less useful than the MSA and will not be considered further.

One such system, for which the MSA has an analytic solution, is the Yukawa fluid. The MSA has been solved for the Yukawa fluid (Waisman, 1973). However, his result is implicit, involving the numerical solution of six simultaneous nonlinear algebraic equations in six unknown variables. Subsequently, (Henderson *et al*, 1995) obtained explicit results for the thermodynamics by expanding the solution in a perturbation expansion. They obtained Eq. (23) for $A_0$ and

$$\frac{A_1}{NkT} = \frac{zL(z)}{12\eta[L(z) + \exp(z)S(z)]}, \tag{41}$$

where $L(z)$ and $S(z)$ are polynomials that are given by

$$L(z) = 12\eta[1 + 2\eta + (1 + \frac{\eta}{2})z] \tag{42}$$

and

$$S(z) = -12\eta(1 + 2\eta) + 18\eta^2 z + 6\eta(1-\eta)z^2 + (1-\eta)^2 z^3. \tag{43}$$

Equation (41) can be obtained by the method of Henderson *et al* but can also be obtained from the PY result for $g_0(R)$ that was derived earlier (Wertheim, 1964).

The expressions for the higher order $A_n$ are complex but are easily calculated. Also it is easy to differentiate the $A_n$ analytically to obtain corresponding expressions for the pressure. Results obtained from the MSA and MC simulation (using $z = 1.8$, which is appropriate for many fluids) are given for $A_1$ through $A_5$ in Figs. 6 and 7. It is virtually impossible to obtain accurate MC values for $A_3 - A_5$ because of the extensive subtraction involved. In an extended vdW theory, Eq. (21) or (28), $A_1$ should be the dominant term. This, certainly, is seen to be the case in Figs. 6 and 7. Also in a vdW theory, $A_1$ should be a linear function of $\rho$. This is not the case but the departure from linearity is not so great. The extended vdW theory is a very reasonable starting point. The simple compressiblity approximation gives similar results for $A_2$. The terms $A_3$ through $A_5$ are always negative (in the MSA, at least) even though it might seem in Fig. 7 that there is a change of sign. In the MSA, the $A_n$ for $n>2$ lack a term that is linear in $\rho$. This is not correct. It is an error in the MSA but is not a serious problem since these higher order terms are small. Additionally, most interest is for dense fluids.

Again it is to be emphasized that if the perturbation series is terminated at $A_1$, then $A_0$, which results from the hard core repulsion, determines the entropy and the contribution of the attractive forces is given by $A_1$. While energy and entropy are mixed through the higher order $A_n$, these higher order terms are small, especially at high densities. The competition between space and attractive forces produces the balance between energy and entropy and minimizes the free energy.

# 5 Real molecules are not hard

In the vdW theory, the modified vdW theory, the MSA, and the perturbation theory discussed so far, the molecules are considered to have hard cores. Real molecules have soft cores. For example, a popular model for a fluid is the Lennard-Jones (LJ) fluid, where the intermolecular potential is given by the LJ 12-6 potential,

$$u(R) = 4\varepsilon\left\{\left(\frac{\sigma}{R}\right)^{12} - \left(\frac{\sigma}{R}\right)^{6}\right\}. \tag{44}$$

The parameter $\varepsilon$ is now the depth of the potential and $\sigma$ is the value of $R$ for which the potential changes sign. The LJ 12-6 potential is plotted in Fig. 8. The LJ 12-6 potential is more reasonable than the Yukawa potential because its repulsive region is not infinitely hard and its long range attraction decays correctly as $R^{-6}$, as is predicted by quantum theory, rather than exponentially. Although an improvement, the LJ 12-6 potential is not a complete representation of the interaction between real molecules. In the case of argon, the depth of the actual interaction is somewhat larger and the the coefficient of the $R^{-6}$ is somewhat smaller than the LJ 12-6 potential suggests. Even so, the Yukawa potential (z=1.8), the LJ 12-6 potential, and more realistic potentials yield similar thermodynamic properties.

Thus, real molecules are not hard. However, the repulsive region is very steep and a hard core representation is very reasonable. In fact, it has been shown (Barker and Henderson, 1967b) that such steep repulsions are accurately represented by a hard sphere potential whose diameter is given by

$$d(T) = \int_0^{\sigma} \{1 - \exp[-\beta u(R)]\}\, dR. \tag{45}$$

The Barker-Henderson diameter, given by Eq. (45), is very weakly temperature dependent. Because of this weak temperature dependence, the separation of energy and entropy terms through first order in the perturbation series is no longer exact. However, this is a small effect.

The RDF of a LJ 12-6 fluid is plotted in Fig. 9 and compared with the hard sphere RDF. It is seen that the hard sphere and LJ RDF's are quite similar. In Fig. 10, the pressure of a LJ 12-6 fluid, calculated from second-order perturbation theory is compared with computer simulations. The agreement is very good.

Subsequently, a useful alternative method for dealing with molecules with a soft core has been developed (Weeks *et al*, 1971a,b)

# 6 Electrolytes

A simple model for electrolytes is the *primitive model* in which the ions are modeled as charged hard spheres and the solvent is modeled as a dielectric continuum, whose dielectric constant is given by $\varepsilon$ (not to be confused with the energy parameter in the intermolecular potentials discussed above). The idea behind the representation of the solvent by a continuum is that the electrolyte is dilute and the ions are far apart and do not sense the molecular nature of the solvent. Thus, the primitive model is useful for low concentrations. The formalism developed so far is applicable to electrolytes except that we should substitute ion for molecule. The ion-ion potential in the primitive model is

$$u(R) = \begin{cases} \infty, & R < \sigma_{ij} \\ \dfrac{z_i z_j e^2}{\varepsilon R}, & R > \sigma_{ij}, \end{cases} \tag{46}$$

where $z_i$ and $\sigma_i$ are the valence and diameter of an ion of species $i$ and $\sigma_{ij} = \sigma_i + \sigma_j$. The parameter $e$ is the magnitude of the electronic charge.

The standard theory for an electrolyte is the DH theory (Debye and Hückel, 1923). Conventionally, this theory is obtained by solving a differential equation, usually linearized, yielding the linearized DH or LDH theory. However, the LDH theory can also be obtained from the MSA or even a chain sum, assuming that the ions are point ions, i.e., $\sigma_i = 0$ (Henderson, 1983). The resulting LDH mean activity coefficient is

$$\ln \gamma_{\pm} = | z_+ z_- | e^2 \frac{\kappa}{2\varepsilon kT}, \tag{47}$$

where

$$\kappa^2 = \frac{4\pi\beta e^2}{\varepsilon} \sum_i z_i^2 \rho. \tag{48}$$

Note that the properties of the electrolyte are typically functions of $\rho^{1/2}$ and/or $T^{1/2}$; this is in contrast to the fluids considered so far, where $\rho$ and $T$ appear with integral powers.

The MSA is the natural generalization of the LDH theory and is obtained by solving the MSA integral or from a chain sum with a nonzero diameter. This was done first by Waisman and Lebowitz (Waisman and Lebowitz, 1970, 1972a,b) for the *restricted primitive model* (RPM), where the ions all have an equal diameter, and by Blum (Blum, 1980) for the PM with ions with differing diameters. At low concentrations the LDH result is valid. That is, the mean activity coefficient, for example, is a linear function of $\kappa$. In contrast, deviations from the LDH result are seen in both theory and experiment as the concentration is increased. However, for variety, the rate constant for an electrochemical reaction can be examined. Reaction rate constants have been calculated using the MSA (Fawcett *et al*, 1997). A typical result is shown in Fig. 11. At low concentrations the rate constant is a linear function $\kappa$, or $I^{1/2}$, where $I$ is the ionic strength, but deviations are seen as the concentration or ionic strength is increased. The MSA result is in good agreement with experiment. The main point is that good results are obtained only when the ion cores are taken into account. In contrast to simple fluids, the ion cores in an electrolyte contribute to both the energy and entropy. The competition is between charge and size but the principle remains.

## 7 Channel selectivity

Ion channels are essential for the proper functions of cells. An ion channel is a protein with a hole through which ions pass selectively. We will consider sodium and calcium channels. Calcium channels play an important role in such physiological functions as muscle contraction. A calcium channel will conduct $Ca^{2+}$ ions when these ions are present in micromolar or larger concentrations even if other ions, say $Na^+$, are present in much larger concentrations whereas the sodium channel does not exhibt such a preference.

A simple intuitive model (Nonner *et al*, 2000), based on the MSA for a bulk electrolyte, in which the selectivity of a calcium channel is produced by the competition between the attractive Coulombic forces between the cations and the negatively charged structural elements, glutamates in what is called a EEEE locus, and the size of the ions and glutamates. A model representation of a calcium channel is shown in Fig. 12, where $\varepsilon_p$ and $\varepsilon_w$ are the delectric coefficients of the protein and the solvent, respectively. The large red spheres in Fig. 12 are the oxygens of the glutamates, each with a single negative charge ($-e$). Since the glutamates are flexible, but attached to the channel protein, the oxygens are confined to the channel but are free to move within the channel. In the EEEE locus there are four glutamates with a negative charge of $-4e$.

In this competition, the ions are attracted into the channel but because of the restricted geometry of the channel filter and the excluded volume of the ions and glutamates, the $Ca^{2+}$ ions are more effective at balancing the $-4e$ charge of the four glutamates than are the $Na^+$ ions since they deliver twice the charge while occupying almost the same volume.

As is seen in Fig. 13, this *charge/size competition* (CSC) model accounts (Boda, *et al*, (2007) very well for the selectivity of a DEEA calcium channel, whose structural elements are (A) aspartate ($-e$), (E) glutamate ($-e$), (A) alanine (neutral), with a net charge of $-3e$ and for the DEKA sodium channel, whose structural elements are aspartate, glutamate, (K) lysine ($+e$), and alanine (called a DEKA locus) with a net charge of $-e$. Because there is little negative charge to balance, the $Na^+$ ions are quite effective at balancing the DEKA charge and the sodium channel is not Ca selective. A simple mutation of K into E changes the net charge from $-e$ to $-3e$ and turns the sodium channel into a calcium channel. Results have been obtained for the EEEE calcium channel (Boda *et al*, 2009).

Besides accounting for Ca/Na selectivity, this model accounts for the selectivity of small ions over large ions of the same valence.

## 8 Summary

It is now well understood that both the molecular size and attractive energy must be taken into account accurately to describe the properties of materials. Examples illustrating this are given from the theory of simple liquids, electrolytes and the selectivity of ion channels. Although this is understood in physical science, it seems less well understood in biological science. For example, the transport of ions is often described by the Nernst-Poisson approach which, like the Debye-Hückel theory, neglects the ion cores. Just as there must be a balance of yin and yang for harmony in life, there must be a balance of size or space or entropy and attractive energy in physical or biological science. It will be interesting to determine by simulation studies the extent to which these concepts are useful in understanding complex systems, such as protein molecules in aqueous solution.

## Acknowledgments

The author is grateful for the support and assistance of his colleagues over the years. Indeed, in many cases the author was the assistant. It is difficult to name all these collaborators. However, Henry Eyring, John Barker, Farid Abraham, Lesser Blum, Ron Fawcett, Andrij Trokhymchuk, Dezsö Boda, Dirk Gillespie, Wolfgang Nonner, and Bob Eisenberg deserve particular mention. The work of the author reviewed here was performed at the CSIRO Chemical Research Laboratories in Melbourne, Australia, the University of Waterloo, Canada, the IBM Research Laboratory in San Jose, California, USA, the University of Utah in Salt Lake City Utah, USA, and Brigham Young University in Provo Utah, USA. The ion channel work was performed using the facilities of the Ira and Marylou Fulton Supercomputing Center at Brigham Young University. The author's trip to Linz was made possible by the support of the Austrian Academy of Science and the University of Linz. Alexis Hales assisted in the preparation of some of the figures. Bob Eisenberg read this manuscript and made useful suggestions.

Figure 1: Pressure-temperature cut of the phase diagram of a typical simple fluid. The solid, liquid, and vapor phases are separated by three coexistence curves that meet at the triple point. The liquid-vapor curve ends with the critical point.

Figure 2: Temperature-density cut of the phase diagram of a typical simple fluid. The lowest temperature is the triple point temperature and the three coexisting phases are marked by circles. The (black) parabolic-like curve gives the liquid-vapor coexistence curve with the critical point, marked by a circle, at the maximum. The region below this curve is a metastable/unstable region. The rapidly rising (red) curves are the solid-fluid coexistence curves. The region between these curves is a metastable/unstable region.

Figure 3: Yukawa intermolecular potential with $z = 1.8$. The distance of closest approach of a pair of particles is $R = \sigma$ at which point the attractive interaction has its greatest magnitude, $u(\sigma) = -\varepsilon$.

Figure 4: Equation of state of a hard sphere fluid. There is a branch for the hard sphere solid at higher densities. The dashed (blue), dot-dashed (red), and solid (black) curves give the results of Eqs. 4, 17, and 18, respectively.

Figure 5: Typical radial distribution function of a dense hard sphere fluid. TheRDF of a Yukawa fluid is similar.

Figure 6: First order perturbation term for a Yukawa fluid with $z = 1.8$. The points give the simulation values of Barker and Henderson while the curve gives the MSA result.

Figure 7: Higher-order perturbation terms for a Yukawa fluid with $z = 1.8$. The curves are labeled with the appropriate value of $b = n$. The points are the simulation values of Barker and Henderson while the curves give the MSA results.

Figure 8: The LJ 12-6 potential. Here $R = \sigma$ is the value for which u(R) changes sign. Since there is no hard core, there is no well defined distance of closest approach. However, a value of $R$ that is somewhat smaller than $\sigma$, determined from Eq. (42), is a reasonable measure of the distance of closest approach.

Figure 9: Typical radial distribution function of a dense LJ 12-6 fluid. The rdf of the unperturbed reference hard sphere fluid, whose diameter is given by Eq. (42), is given by the broken curve.

Figure 10: Equation of state of the LJ 12-6 fluid. The curves are isotherms whose corresponding values of $kT/\varepsilon$ are shown and the points are computer simulation values for the same isotherms. The temperatures $kT/\varepsilon$=1.35 and 0.72 are approximately the critical and triple point temperatures.

Figure 11: Values of the rate constant for the system $BrCH_2COO^-$-$S_2O_3^{-2}$ as a function of the ionic strength. The broken straight line is the LDH result whereas the solid curve is the MSA result.

Figure 12: Model calcium (or sodium) channel filter. The red spheres represent the amino acids that are tethered to the channel protein, which has a dielectric coefficient, $\varepsilon_p$, that is lower than that of the electrolyte, $\varepsilon_w = 80$. The other spheres in the filter are the ions.

Figure 13: Computer simulation results for the occupancy of a DEEA calcium channel with a net charge of -3$e$ and a DEKA sodium channel with a net charge of -$e$. In each case, $Ca^{2+}$ ions are added to a 0.1M NaCl solution in the reservoir. The DEEA channel is calcium selective because even a small addition of $Ca^{2+}$ ions excludes $Na^{+}$ ions from the filter.

Fig 1

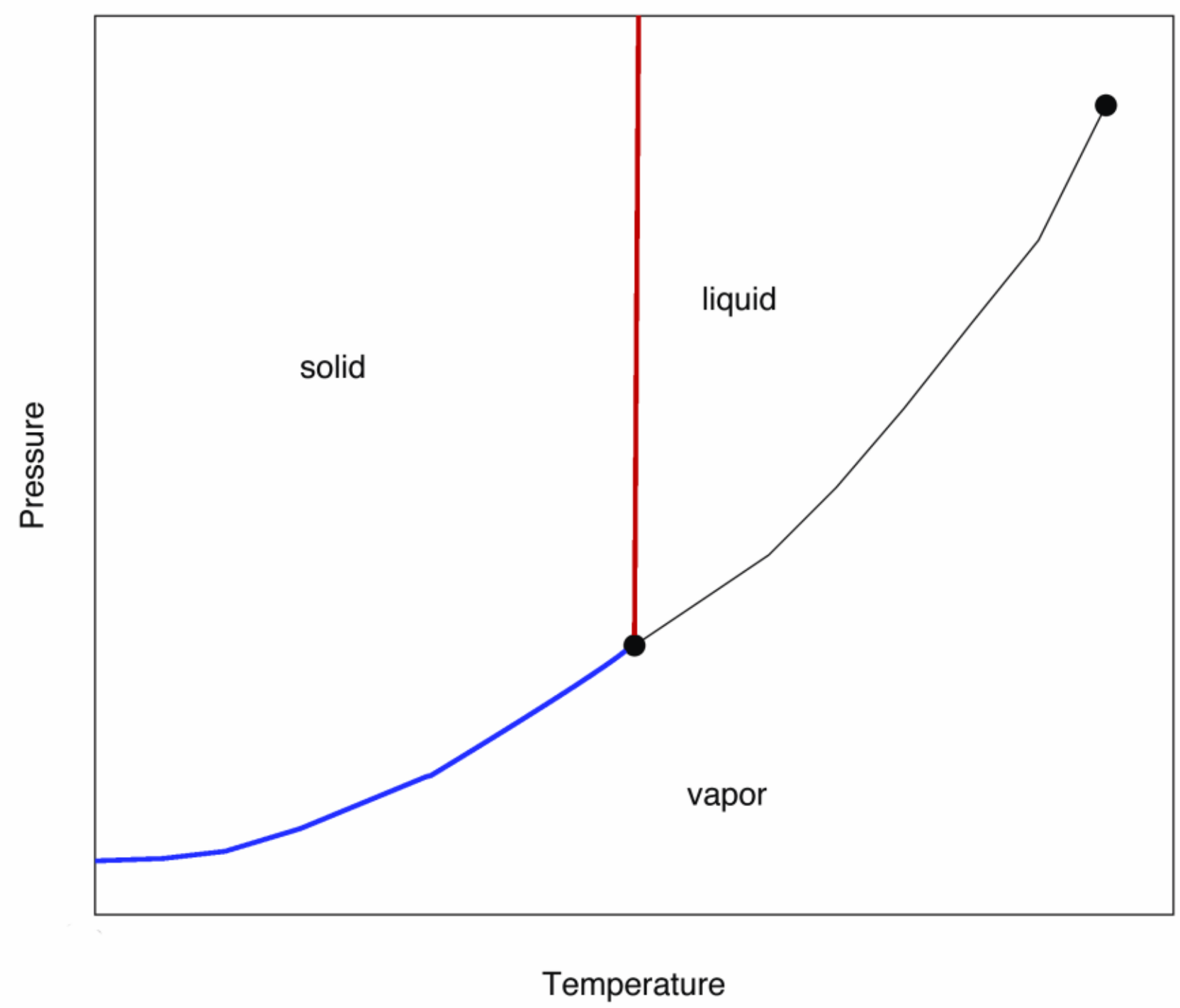

Fig 2

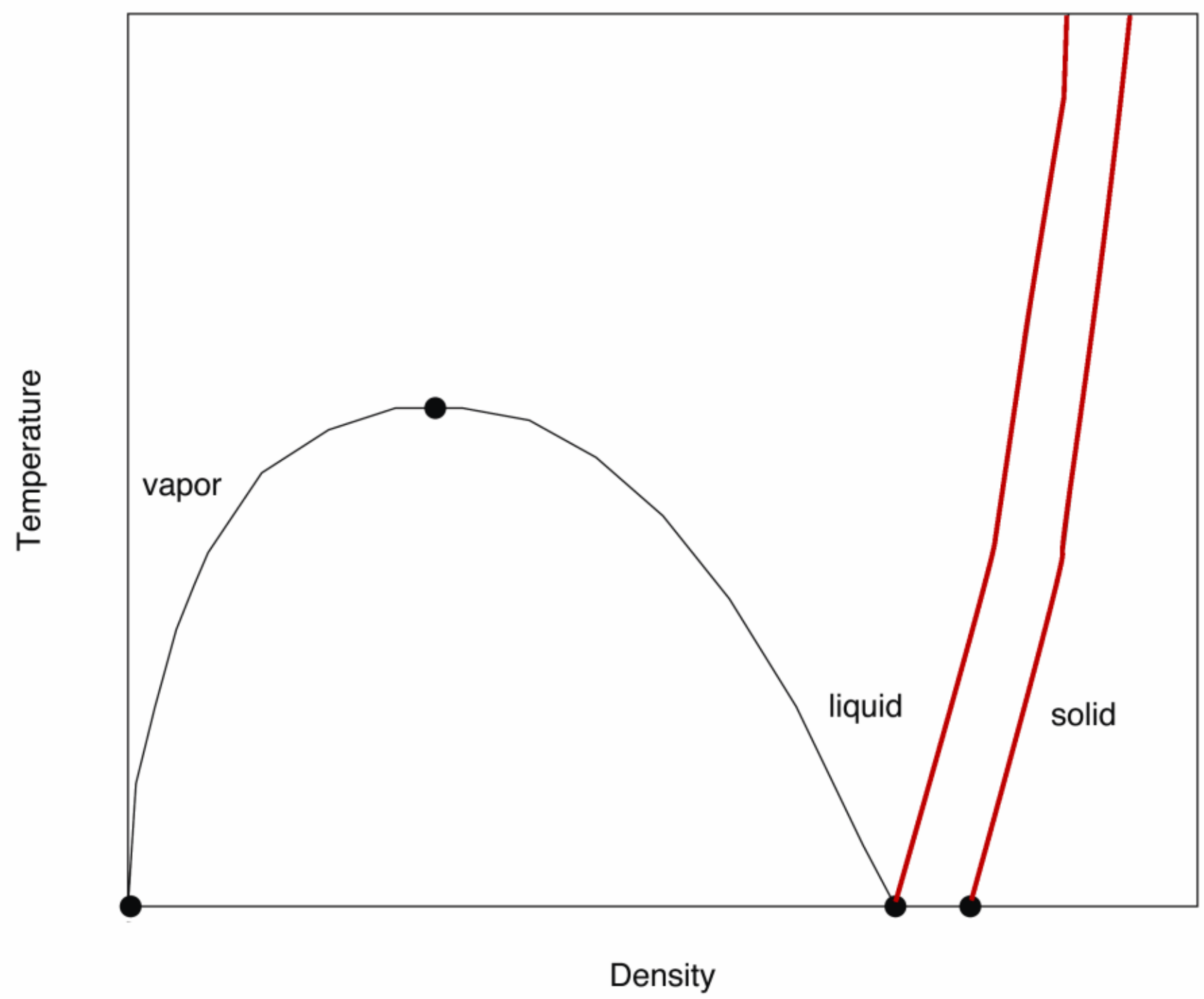

Fig 3

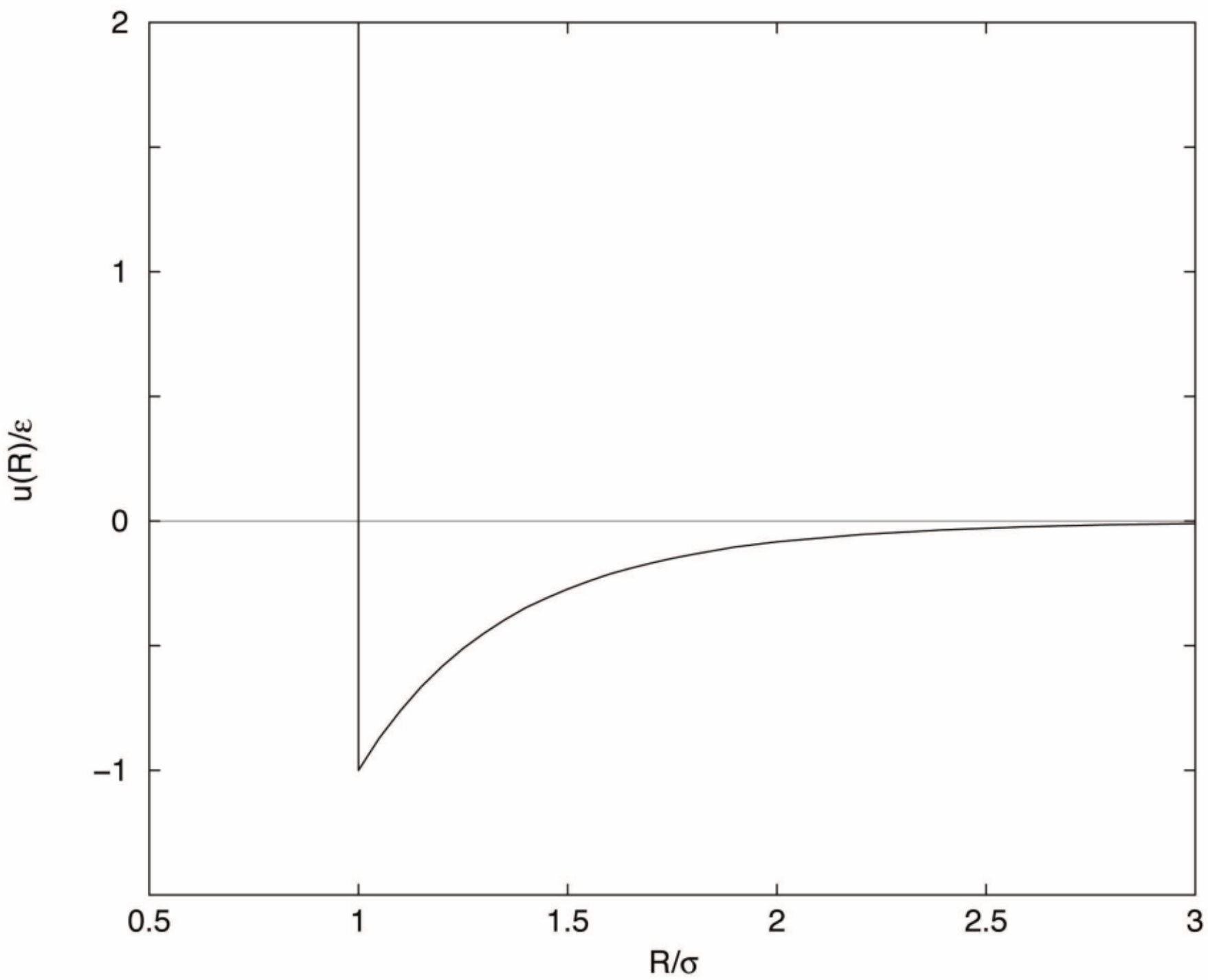

Fig 4

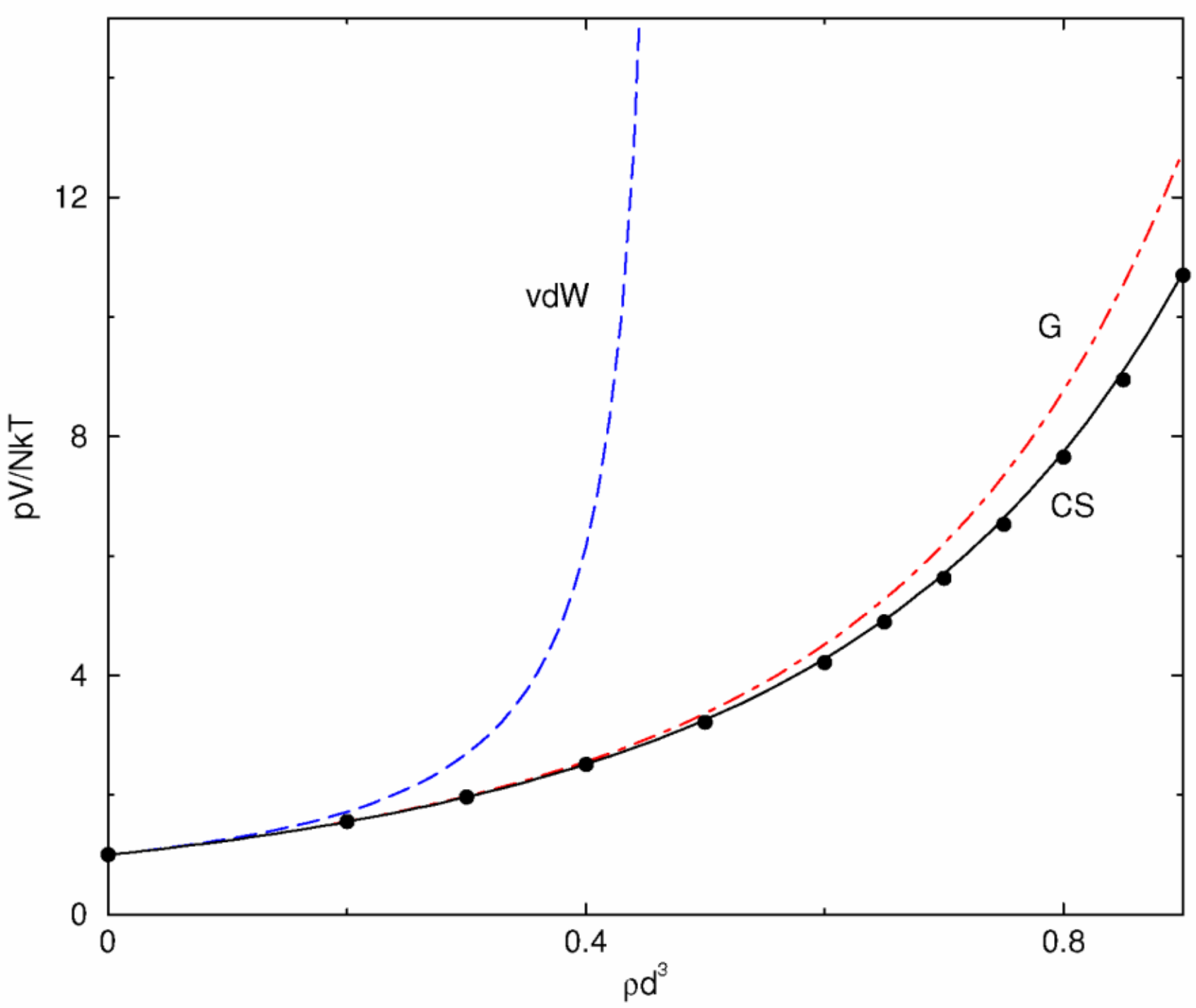

Fig 5

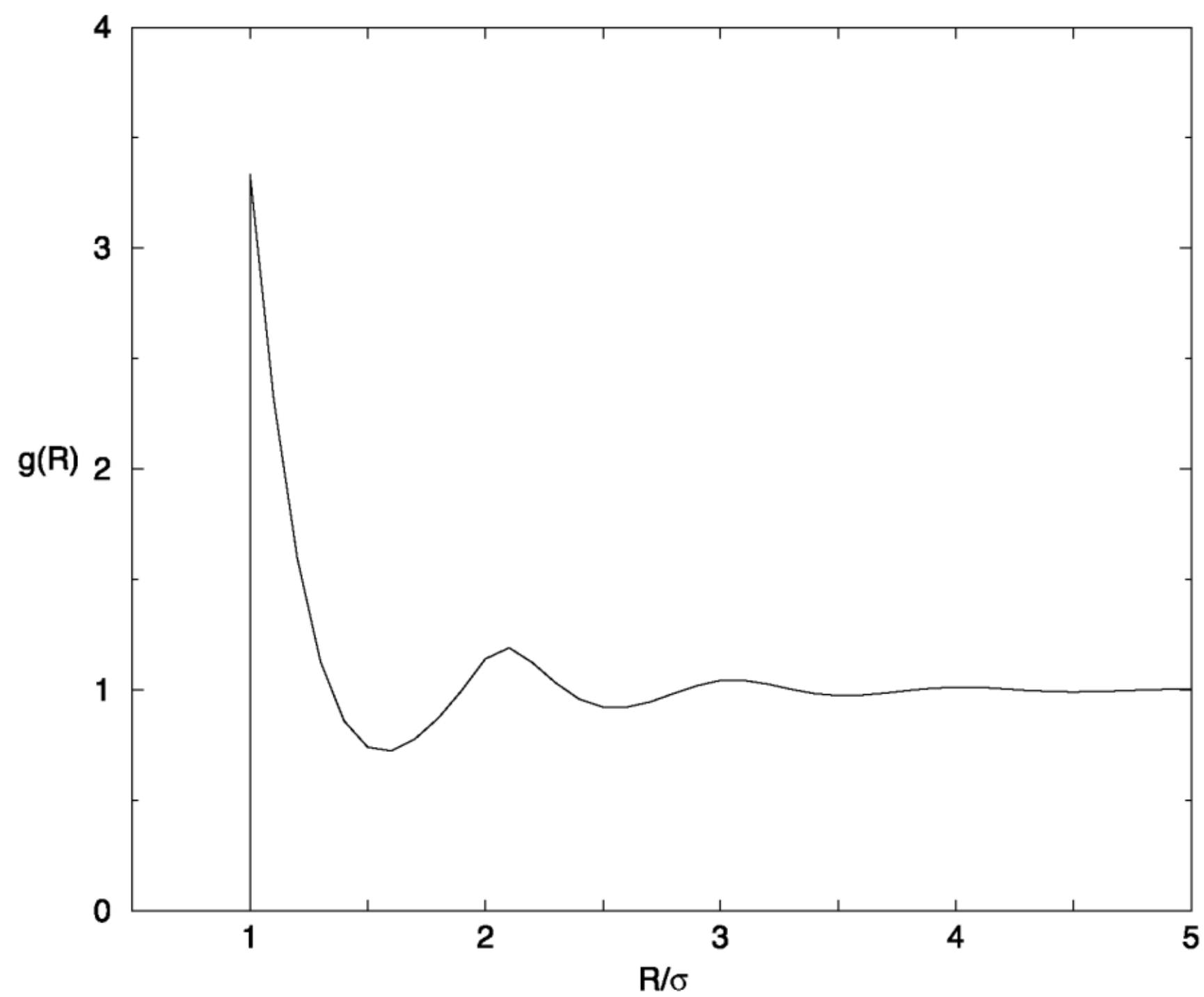

Fig 6

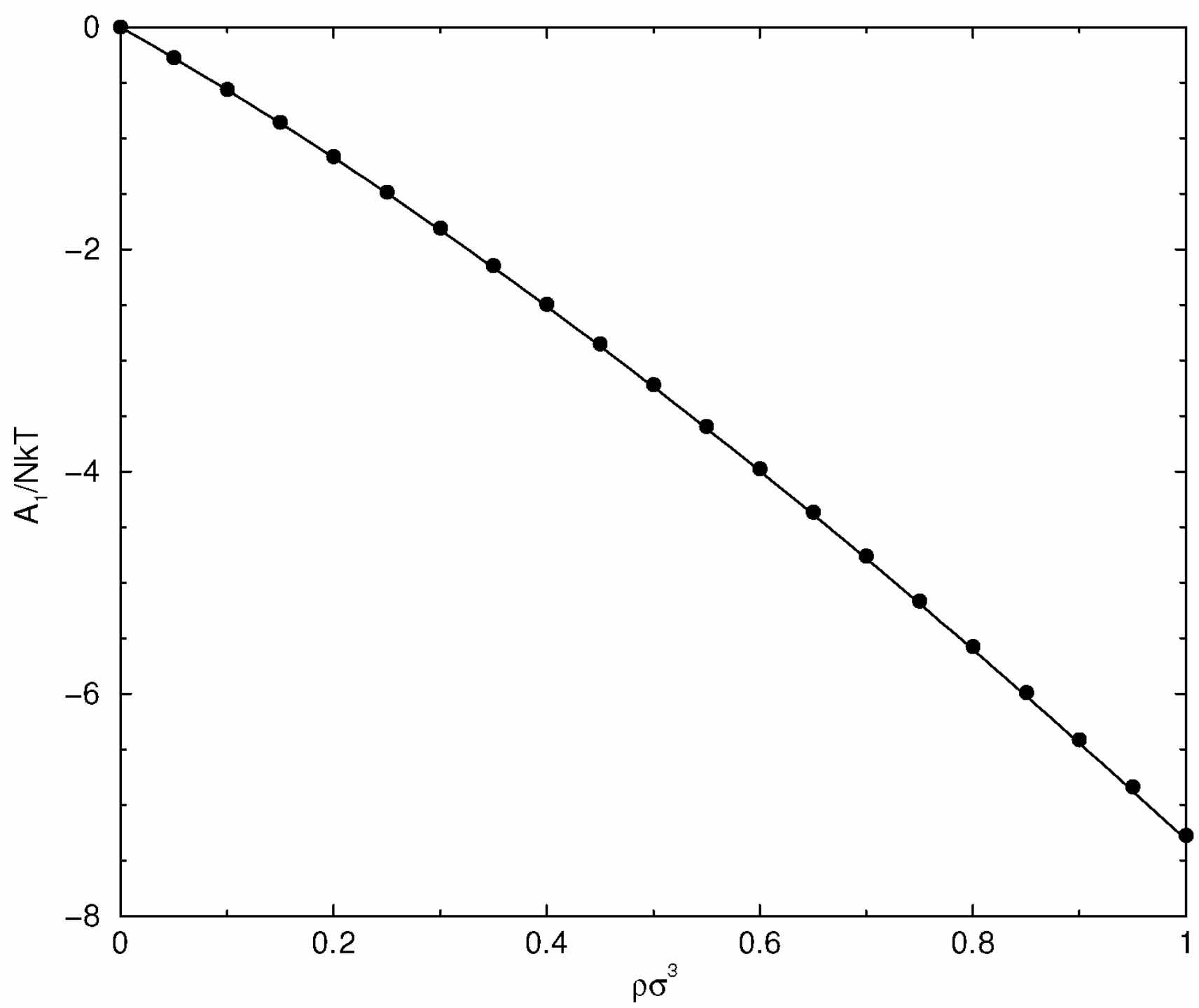

Fig 7

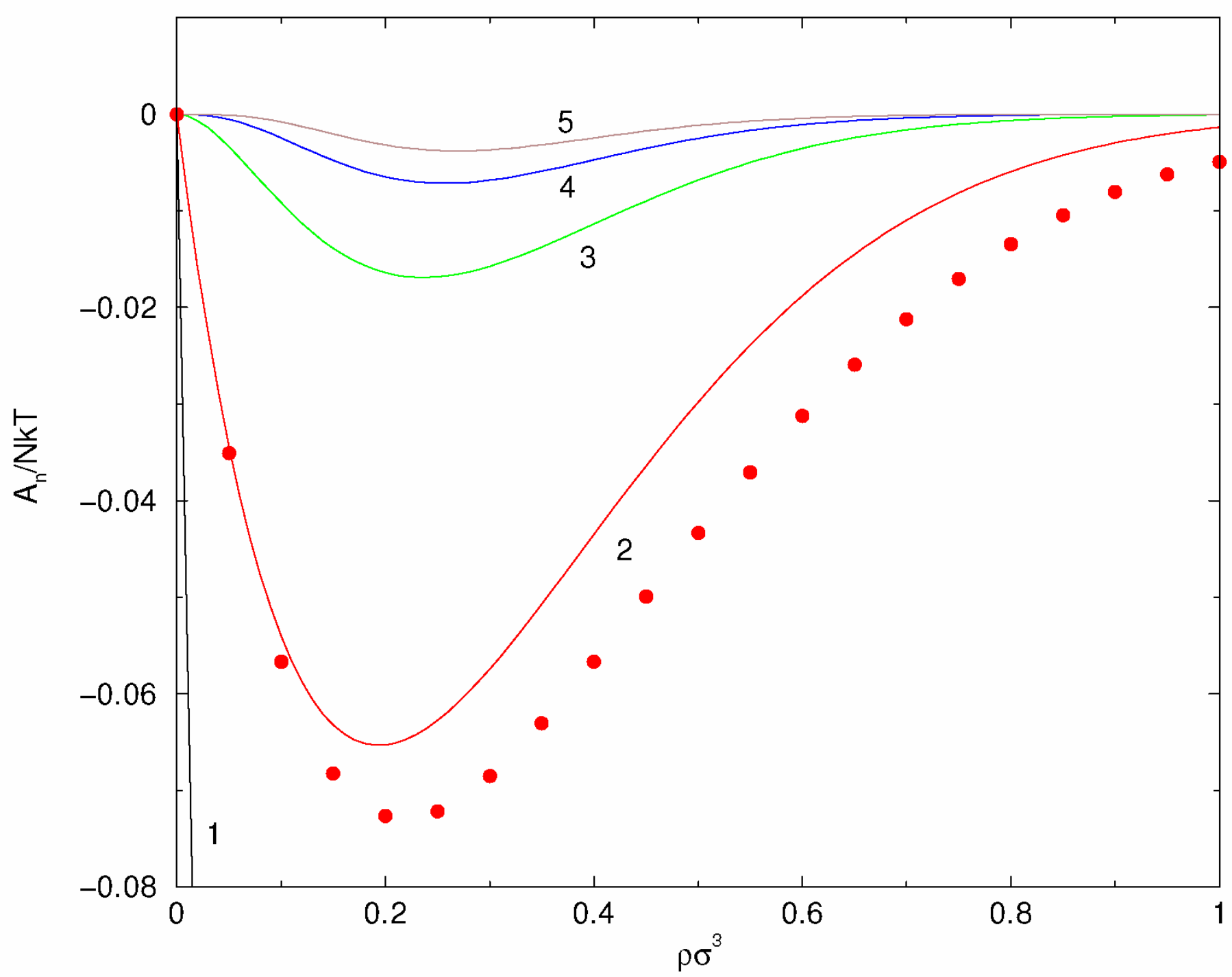

Fig 8

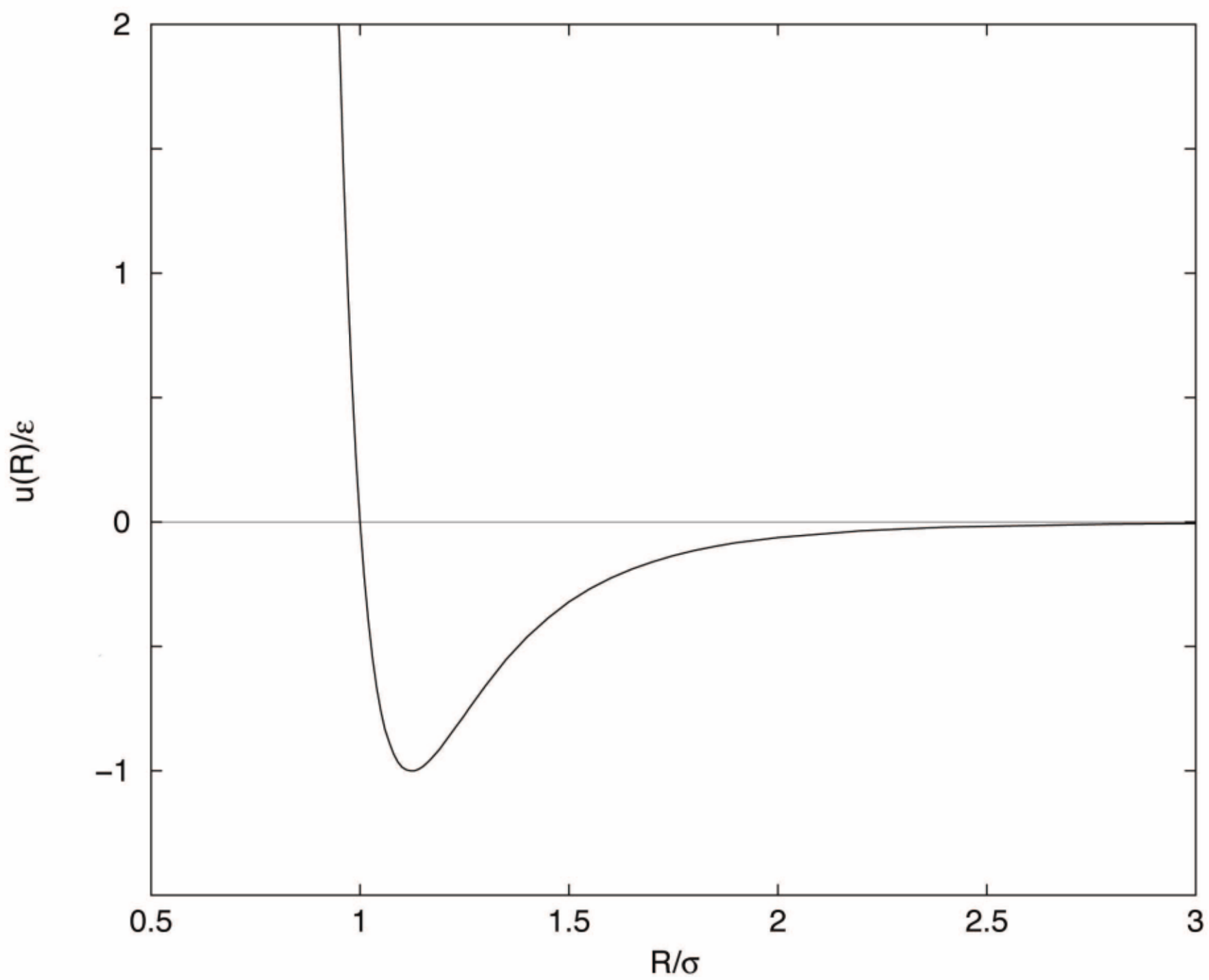

Fig 9

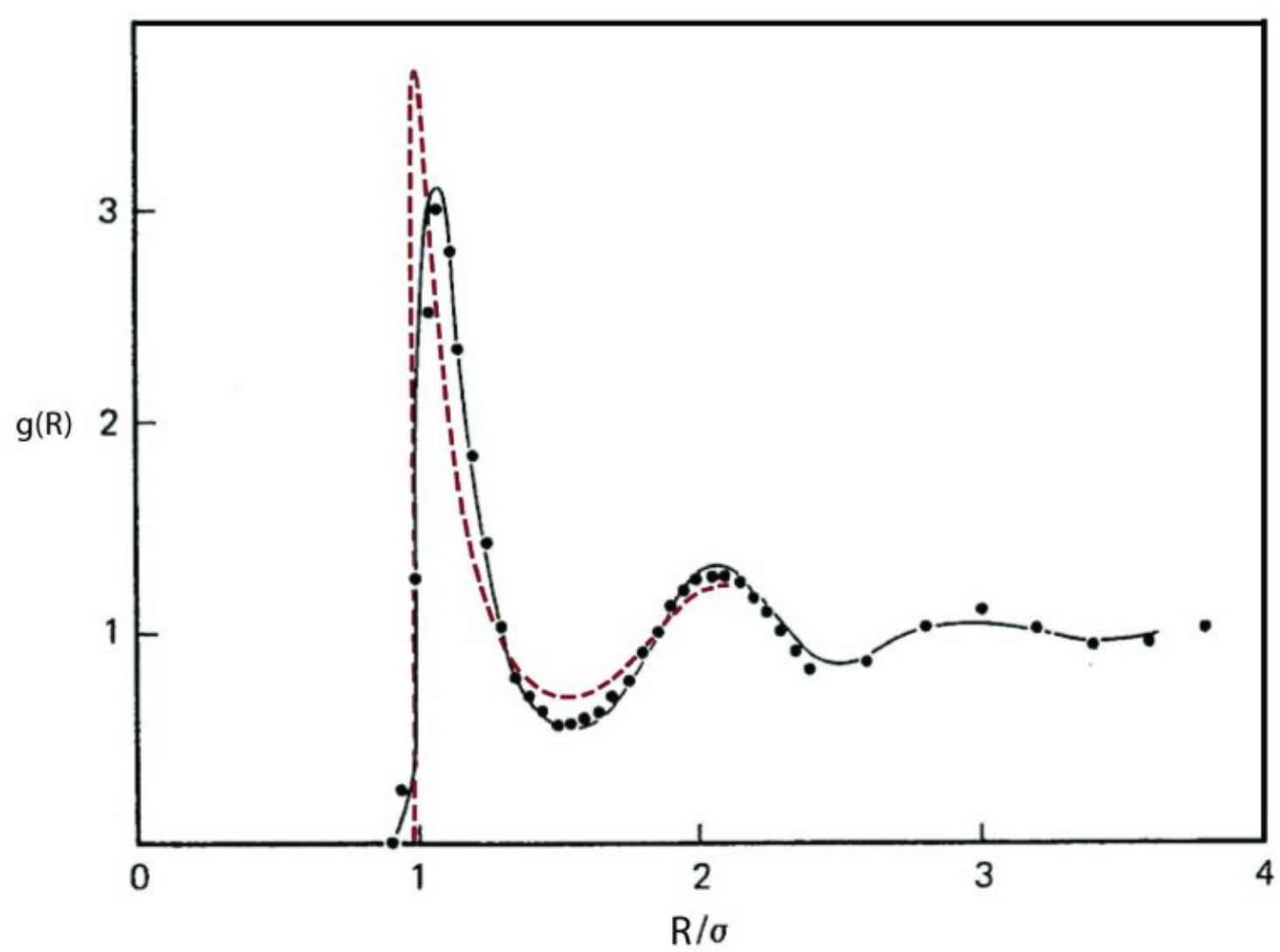

Fig 10

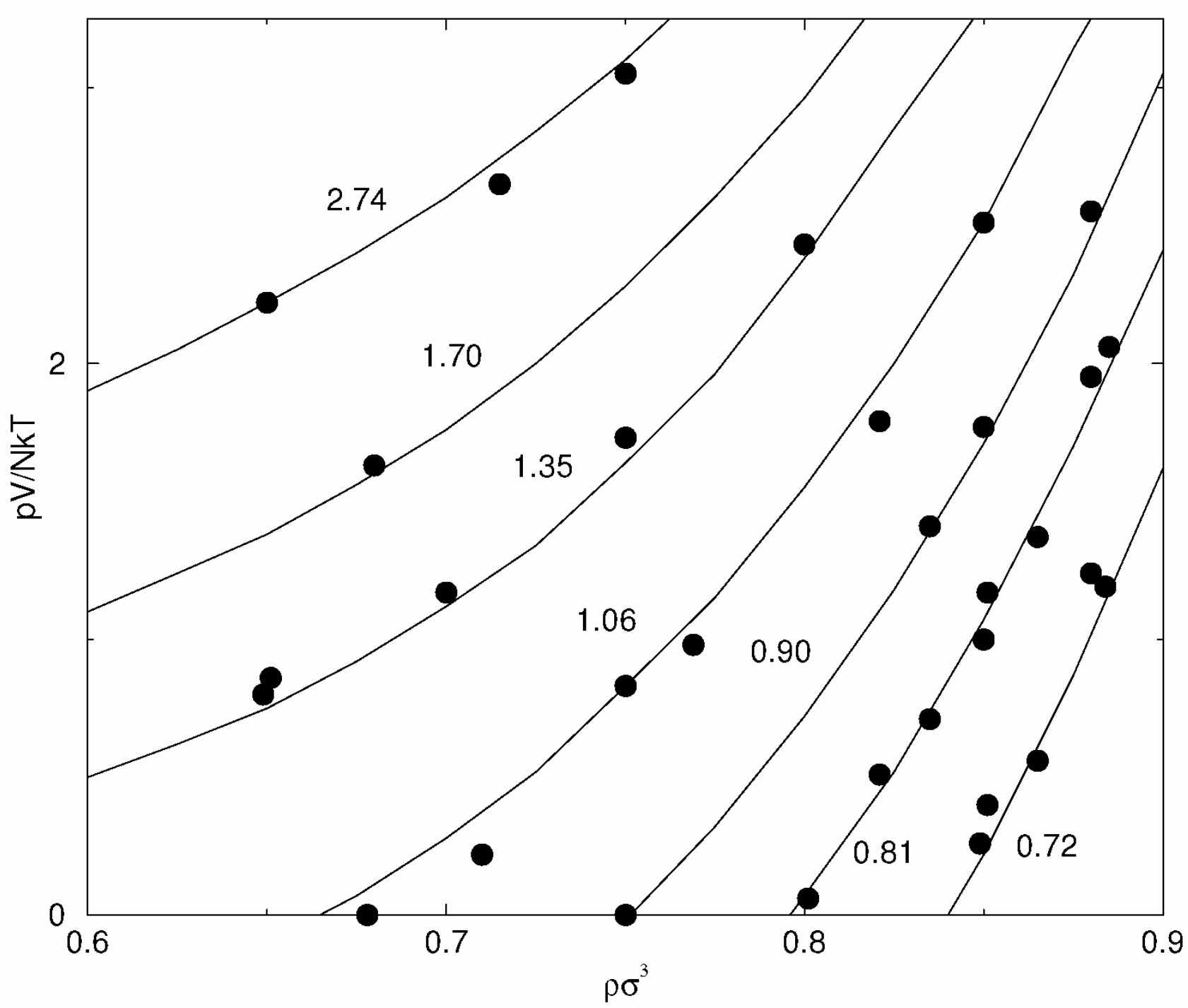

Fig 11

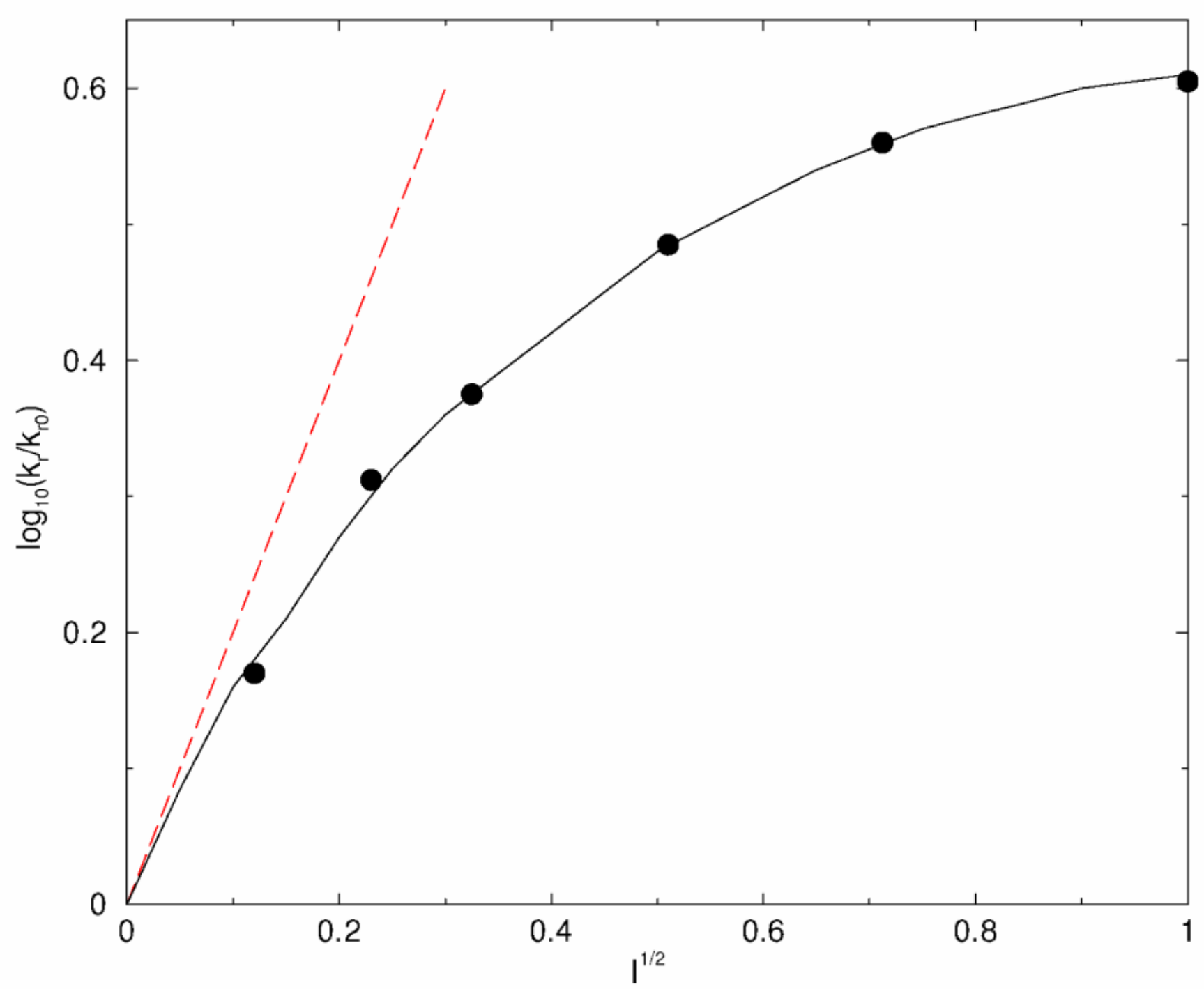

Fig 12
TOO LARGE FOR arXiv

Fig 13

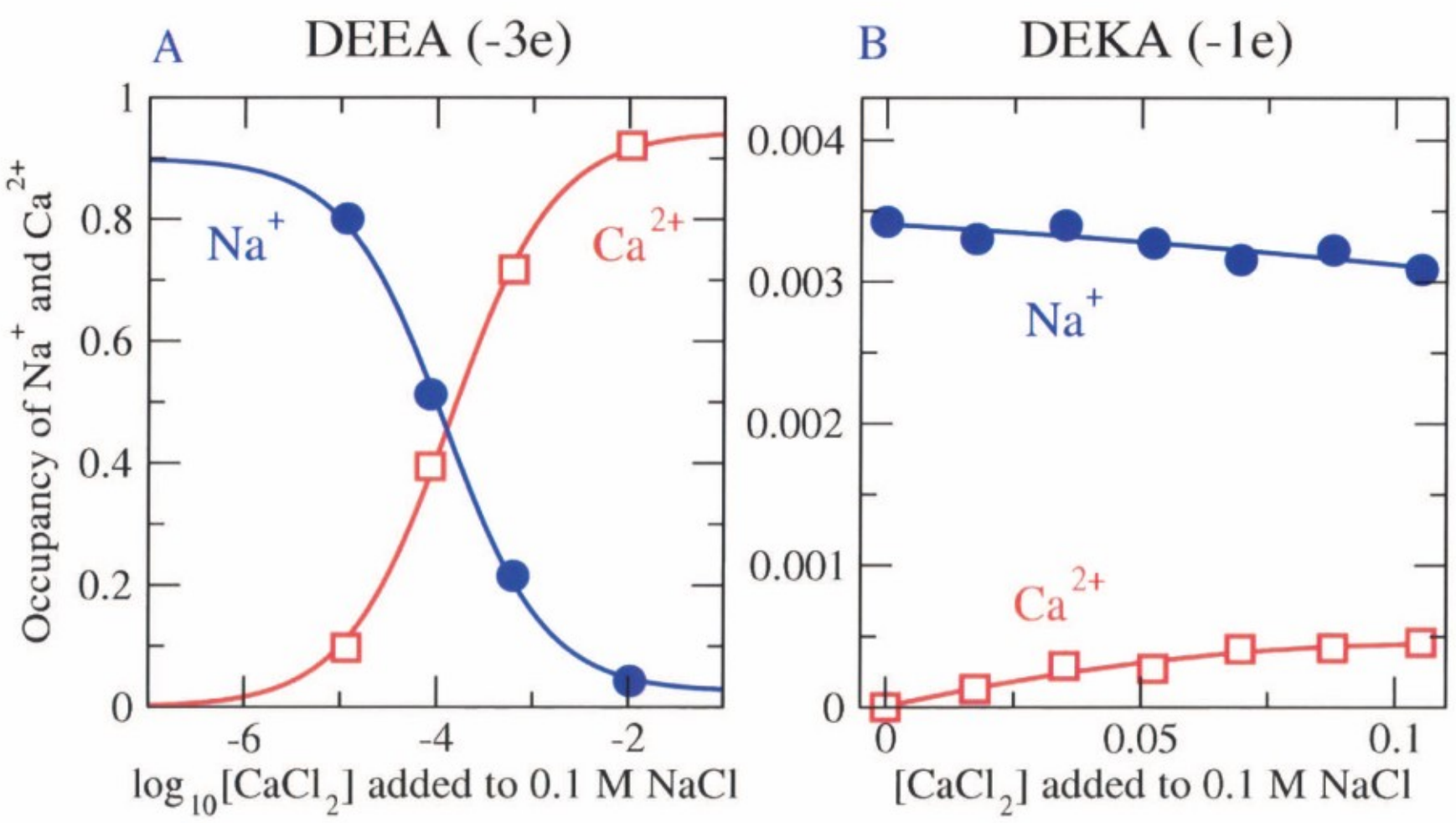
A
DEEA (-3e)
B
DEKA (-1e)
Occupancy of Na+ and Ca2+
Na+
Ca2+
Na+
Ca2+
1
0.8
0.6
0.4
0.2
0
-6
-4
-2
log10[CaCl2] added to 0.1 M NaCl
0.004
0.003
0.002
0.001
0
0.05
0.1
[CaCl2] added to 0.1 M NaCl